\documentclass[11pt,a4paper]{article}

\usepackage{amsmath}
\usepackage{amssymb}
\usepackage{graphicx}
\usepackage{url}
\usepackage{epsfig}
\usepackage[T1]{fontenc}
\usepackage[latin9]{inputenc}
\usepackage{multirow}
\newcommand{\tb}{\bar t}

\newcommand{\ttbh}{ t \tb H}
\newcommand{\al}{\alpha}
\newcommand{\als}{\alpha_{\rm s}}
\newcommand{\shat}{\hat s}

\newcommand{\muf}{\mu_{\rm F}}
\newcommand{\mur}{\mu_{\rm R}}
\newcommand{\mufo}{\mu_{{\rm F},0}}
\newcommand{\muro}{\mu_{{\rm R},0}}
\newcommand{\sigh}{\hat \sigma}

\newcommand{\nn}{\nonumber}

\newcommand{\tosv}{{\scriptscriptstyle \to}}

\RequirePackage[numbers,sort&compress]{natbib}

\def\beq{\begin{equation}}
\def\eeq{\end{equation}}
\def\bear{\begin{eqnarray}}
\def\eear{\end{eqnarray}}

\def\bet34{\beta_{kl}}
\def\CF{C_{\mathrm{F}}}
\def\CA{C_{\mathrm{A}}}
\def\nf{n_{\mathrm{f}}}
\def\TR{T_{\mathrm{R}}}
\begin{document}

\begin{flushright}
	MS-TP-17-06
\end{flushright}
\vspace{1cm}

\begin{center}
	{\Large \textbf{Associated $t \tb H$ production at the LHC: theoretical predictions at NLO+NNLL accuracy}}\\
	\vspace{.7cm}
	Anna Kulesza$^{a,}$\footnote{\texttt{anna.kulesza@uni-muenster.de}}, Leszek Motyka$^{b,}$\footnote{\texttt{leszekm@th.if.uj.edu.pl}}, Tomasz Stebel$^{b,c,}$\footnote{\texttt{tomasz.stebel@uj.edu.pl}} and Vincent Theeuwes$^{d,}$\footnote{\texttt{vtheeuwe@buffalo.edu}}

	\vspace{.3cm}
	\textit{
		$^a$ Institute for Theoretical Physics, WWU M\"unster, D-48149 M\"unster, Germany\\
		$^b$ Institute of Physics, Jagellonian University, S.\L{}ojasiewicza 11, 30-348 Krak\'ow, Poland\\
        $^c$ Institute of Nuclear Physics PAN, Radzikowskiego 152, 31-342 Krak\'ow, Poland\\
		$^d$ Department of Physics, SUNY Buffalo, 261 Fronczak Hall, Buffalo, NY 14260-1500, USA
	}
\end{center}   

\vspace*{2cm}
\begin{abstract}
We perform threshold resummation of soft gluon corrections to the total cross section and the invariant mass distribution for the process $pp \to \ttbh$. The resummation is carried out at next-to-next-to-leading-logarithmic (NNLL) accuracy using the direct QCD Mellin space technique in the three-particle invariant mass kinematics. After presenting analytical expressions we discuss the impact of resummation on the numerical predictions for the associated Higgs boson production with top quarks at the LHC. We find that NLO+NNLL resummation leads to predictions for which the central values are remarkably stable with respect to scale variation and for which theoretical uncertainties are reduced in comparison to NLO predictions.
\end{abstract}


\clearpage
\tableofcontents
\setcounter{footnote}{0}

\section{Introduction}
\label{s:intro}

The measurement of Higgs boson production rates in the $pp \to \ttbh$ process  is of the central importance to the LHC research program. The process has been intensively searched for in Run~1~\cite{Aad:2014lma,Aad:2015gra,Khachatryan:2014qaa,Khachatryan:2015ila,Aad:2015iha} and its measurement  is among the highest priorities of the LHC Run~2 physics programme~\cite
{deFlorian:2016spz,ATLAS:2016ldo,CMS:2016rnk}. 
The associated production process offers a direct way to probe the strength of the top--Higgs Yukawa coupling without making any assumptions regarding its nature. As the top--Higgs Yukawa coupling is especially sensitive to the underlying physics, $\ttbh$ production provides a vital test of the Standard Model (SM) and possibly a means to probe the beyond the SM physics indirectly. It is thus highly important that precise and reliable theoretical predictions are available for this process.

For these reasons, a large amount of effort has been invested to improve theoretical description of the $\ttbh$ production. While the next-to-leading-order (NLO) QCD, i.e.\ ${\cal O}(\als^3\al)$ predictions were obtained some time ago~\cite{Beenakker:2001rj, Reina:2001sf}, they have been newly recalculated and matched to parton showers in~\cite{Hirschi:2011pa, Frederix:2011zi, Garzelli:2011vp, Hartanto:2015uka}. In the last years, the mixed QCD-weak corrections~\cite{Frixione:2014qaa} and QCD-EW corrections~\cite{Yu:2014cka, Frixione:2015zaa} of ${\cal O}(\als^2\al^2)$ are also available. Furthermore, the NLO EW and QCD corrections to the hadronic $\ttbh$ production with off-shell top and antitop quarks  have been recently obtained~\cite{Denner:2015yca, Denner:2016wet}. For the most part, the NLO QCD corrections are $\sim20$\% at the Run 2 LHC energies, whereas the size of the (electro)weak correction is more than ten times smaller. The scale uncertainty of the NLO QCD corrections is estimated to be $\sim10$\%~\cite{Beenakker:2001rj, Reina:2001sf, Dittmaier:2011ti}. 

In general, if for a given process one expects that a significant part of higher order corrections originates from emission of soft and/or collinear gluons, it is possible to improve the accuracy of theoretical predictions by employing methods of resummation. Relying on principles of factorization between various dynamical modes, they allow an all-order calculation of dominant logarithmic corrections originating from a certain  kinematical limit.  Supplementing fixed-order results with resummation leads not only to change in the value of the cross section but also offers a better control over the theoretical error, in particular due to cancellations of factorization scale dependence between parton distribution functions (pdfs) and the partonic cross sections. The universality of resummation concepts warrants their applications to scattering processes with arbitrary many partons in the final state~\cite{Bonciani:2003nt, Aybat:2006wq}, thus also to a class $2 \to  3$ processes and in particular the associated $\ttbh$ production at the LHC.

The first step in this direction was performed by us in~\cite{Kulesza:2015vda}, where we presented the first calculation of the resummed total cross section for the $\ttbh$ production at the next-to-leading-logarithmic (NLL) accuracy. The calculation relied on application of the traditional Mellin-space resummation formalism in the absolute threshold limit, i.e.\ in the limit of the partonic energy $\sqrt{\shat}$ approaching the production threshold $M=2 m_t + m_H$, $\shat \to M^2$, where $m_t$ is the top quark mass and $m_H$ is the Higgs boson mass. In~\cite{Kulesza:2015vda}, we have achieved an all-order improvement of the theoretical predictions by taking into account a well-defined subclass of higher order corrections.   However, due to the suppression of the available 3-particle phase-space in the absolute production threshold limit, it is to be expected that the numerical impact of formally large logarithmic corrections resummed in these kinematics will be somewhat diminished and that contributions prevailing numerically might come from regions further away from the absolute threshold scale $M$. 

Subsequently we have performed~\cite{Kulesza:2016vnq}  resummation of NLL corrections arising in the limit of $\sqrt {\shat}$ approaching the invariant mass threshold $Q$ with $Q^2= (p_t +p_{\bar t}+ p_H)^2$. We have considered  cross sections differential in the invariant mass $Q$, as well as the total cross sections obtained after integration over $Q$.  For a $2 \to 2$ process, this type of resummation is often referred to as threshold resummation in the pair-invariant mass (PIM) kinematics. Threshold resummation can be also performed in the framework of the soft-collinear effective theory (SCET). The first application of this technique to a $2 \to 3$ process, more specifically to the process $pp \to t\bar t W^{\pm}$, was presented in~\cite{Li:2014ula}. The SCET framework was also used to obtain an approximation of the next-to-next-to-leading order (NNLO) $\ttbh$ cross section and distributions~\cite{Broggio:2015lya}, following from the expansion of the NNLL resummation formula. Recently, NNLL results for $t \bar t W$~\cite{Broggio:2016zgg}, $\ttbh$~\cite{Broggio:2016lfj} and $t \bar t Z$~\cite{Broggio:2017kzi} associated production processes appeared, based on expressing the SCET formulas in Mellin space.

In this paper, we continue the work presented in~\cite{Kulesza:2016vnq} and perform threshold resummation in the invariant mass limit at the NNLL accuracy using the direct QCD~\cite{Sterman:2013nya} Mellin-space approach.
Compared to NLL calculations,  the anomalous dimensions governing resummation need to be implemented with accuracy higher by one order. In contrast to the absolute threshold limit considered in~\cite{Kulesza:2015vda}, the soft anomalous dimension is a matrix in the colour space containing non-zero off-diagonal elements, thus requiring an implementation of the diagonalization procedure. We then match our NNLL cross section with the fixed-order cross section at NLO. The invariant mass kinematics also offers an opportunity to perform resummation for the differential distributions in $Q$. 

The paper is structured as follows. In Section 2 we review threshold resummation in Mellin space, stressing the difference between the resummation in the invariant mass and the absolute threshold limits. The numerical results and their discussion is presented in Section 3, where we also compare our results to those in~\cite{Broggio:2016lfj}.  We summarize our most important findings in Section 4.

\section{NNLL resummation in the triple invariant mass kinematics for $2 \to 3$ processes with two massive coloured particles in the final state}
\label{s:theory}

The resummation of soft gluon corrections to the differential cross section $d\sigma_{pp \to \ttbh}/ dQ^2$ is performed in Mellin space, where the Mellin moments are taken w.r.t.\ the variable \mbox{$\rho = Q^2/S$}. At the partonic level, the Mellin moments for the process $ ij \to kl B$, where $i,j$ denote massless coloured partons, $k, l$ two massive quarks and $B$ a massive colour-singlet particle, are given by
\beq
\frac{d \tilde \sigh_{ij \to kl B} }{d Q^2}(N,Q^2,\{m^2\},\muf^2, \mur^2) = \int_0^1 d \hat\rho \, \hat\rho^{N-1} \frac{ d\sigh_{ij \to kl B}}{d Q^2} (\hat \rho, Q^2,\{m^2\},\muf^2, \mur^2) ,
\eeq
with $\hat \rho = Q^2/\shat$ and $\{m^2\}$ denoting all masses entering the calculations.

Taking the Mellin transform allows one to systematically treat the logarithmic terms of the form $\als^n \left[ \log^m (1-z)/(1-z)\right]_+$, with $m\leq 2n-1$ and $z=Q^2/\shat$, appearing in the perturbative expansion of the partonic cross section to all orders in $\als$. In Mellin space these logarithms turn into logarithms of the variable $N$, and the threshold limit $z \to 1$ corresponds to the limit $N \to \infty$.

The resummed cross section in the $N$-space has the form~\cite{Contopanagos:1996nh,Kidonakis:1998nf}\footnote{In fact, the soft function $\mathbf{S}_{ij\to klB}$ as well as the radiative factors $\Delta^i$, $\Delta^j$ are dimensionless  functions of the ratios of the scales and the coupling constant at the renormalization scale. The current notation indicating dependence on the scales is introduced for brevity.} 
\bear
\label{eq:res:fact_master}
\frac{d \tilde\sigh^{{\rm (res)}}_{ij\tosv kl B}}{dQ^2}&&\hspace{-0.9cm}(N,Q^2,\{m^2\},\muf^2, \mur^2) = \\ \nn
&=& {\mathrm{Tr}} \left[ \,\mathbf{H}_{ij\tosv kl B} (Q^2,\{m^2\},\muf^2, \mur^2) 
 \mathbf{S}_{ij\to klB}(N+1,Q^2,\{m^2\},\mur^2) \,  \right] \, \\ \nn
&\times& \Delta^i(N+1,Q^2,\muf^2, \mur^2) \Delta^j(N+1,Q^2,\muf^2, \mur^2),
\eear
where the trace is taken over colour space. The appearance of colour dependence in Eq.~(\ref{eq:res:fact_master}) is inherently related to the fact that  soft radiation is coherently sensitive to the colour structure of the hard process from which it is emitted. The matrix $\mathbf{H}_{ ij\tosv kl B}$ indicates the hard-scattering contributions, absorbing the off-shell effects, projected onto the chosen colour basis. The colour matrix $\mathbf{S}_{ij\to klB}$ represents the soft wide-angle emission.  The functions $\Delta^i$ and $\Delta^j$ sum the logarithmic contributions due to (soft-)collinear radiation from the incoming partons.  The radiative factors are thus universal for a specific initial state parton, i.e.\ they depend neither on the underlying colour structure nor the process.

At LO the $\ttbh$ production receives contributions from the $q \bar q$ and $gg$ channels. We analyze the colour structure of the underlying processes in the $s$-channel colour bases, $\{ c_I^q\}$ and $\{c_I^g\}$, with 
$$c_{\bf 1}^q =  \delta^{\al_{i}\al_{j}} \delta^{\al_{k}\al_{l}}, \quad c_{\bf 8}^{q} = T^a_{\al_{i}\al_{j}} T^a_{\al_{k}\al_{l}},
$$
$$c_{\bf 1}^{g} = \delta^{a_i a_j} \, \delta^{\al_k
  \al_l}, \quad c_{\bf 8S}^{g}=  T^b _{\alpha_l \alpha_k} d^{b a_i a_j} ,\quad
c^{g}_{\bf 8A} = i T^b _{\alpha_l \alpha_k} f^{b a_i a_j} .
$$ 

The hard function $\mathbf{H}_{ ij\tosv kl B}$ carries no dependence on $N$ and is given by the perturbative expansion
\beq
\mathbf{H}_{ ij\tosv kl B}= \mathbf{H}^{\mathrm{(0)}}_{ij\to klB} + \frac{\als}{\pi}\mathbf{H}^{\mathrm{(1)}}_{ij\to klB} + \dots
\label{eq:hard:nlo}
\eeq
In order to perform resummation at NLL accuracy one needs to know $\mathbf{H}^{\mathrm{(0)}}_{ij\to klB}$, whereas NNLL accuracy requires the knowledge of the $\mathbf{H}^{\mathrm{(1)}}_{ij\to klB}$ coefficient.

The soft function, on the other hand,  resums logarithms of $N$ at the rate of one power of the logarithm per  power of the strong coupling. These single logarithms due to the soft emission can be confronted with double logarithms due to soft and collinear emissions resummed by the jet factors $\Delta^i$ and $\Delta^j$. As a dimensionless function, $\mathbf{S}_{ij\to klB}$ depends only on the ratio of the scales. At the same time, the dependence on $N$ enters only via $Q/N$~\cite{Czakon:2009zw}, making $\mathbf{S}_{ij\to klB}$ dependent on the ratio $Q/(N\mur)$. The soft function is given by a solution of the renormalization group equation~\cite{KS2,Kidonakis:1998nf} and has the form
\bear
 \mathbf{S}_{ij\to klB}(N,Q^2,\{m^2\}, \mur^2)&= &\mathbf{\bar{U}}_{ij\tosv kl B}(N, Q^2,\{m^2\}, \mur^2 )\  \mathbf{\tilde S}_{ij\to klB}(\als(Q^2/{\bar N^2})) \nn \\ 
&\times &\mathbf{{U}}_{ij\tosv kl B}(N,Q^2,\{m^2\}, \mur^2), 
\label{eq:soft:evol}
\eear
where $\mathbf{\tilde S}_{ij\to klB}$ plays a role of a boundary condition and is obtained by taking $\mathbf{S}_{ij\to klB}$ at \mbox{$Q^2/(\bar N^2\mur^2)=1$} with $\bar{N}=Ne^{\gamma_{\rm E}}$ and ${\gamma_{\rm E}}$ denoting the Euler constant. It is a purely eikonal function~\cite{KS2,Kidonakis:1998nf,Dixon:2008gr} and can be calculated perturbatively 
\beq
\mathbf{\tilde S}_{ij\to klB}=\mathbf{\tilde S}^{\mathrm{(0)}}_{ij\to klB} + \frac{\als}{\pi}\mathbf{\tilde S}^{\mathrm{(1)}}_{ij\to klB} + \dots
\label{eq:soft:nlo}
\eeq
At the lowest order the colour matrix is given  by:
\begin{equation}
\left(\mathbf{\tilde S}^{\mathrm{(0)}}_{ij\to kl}\right)_{IJ}=\mathrm{Tr}\left[c_I^\dagger c_J\right].
\end{equation}
Similarly to the hard function,  knowledge of $\mathbf{S}^{\mathrm{(0)}}_{ij\to klB}$ is required in order to perform resummation at NLL accuracy and a result for $\mathbf{S}^{\mathrm{(1)}}_{ij\to klB}$ at the NNLL accuracy.
The scale of $\als$ in $\mathbf{\tilde S}_{ij\to kl}$, equal to $Q^2/\bar{N}^2$, results in an order $\als^2\log\bar{N}$ term if we expand $\mathbf{\tilde S}_{ij\to kl}$ in $\als (\mur^2)$. 

The soft function evolution matrices $\mathbf{\bar{U}}_{ij\tosv kl B}$, $\mathbf{{U}}_{ij\tosv kl B}$ contain  logarithmic enhancements due to soft wide-angle emissions~\cite{Kidonakis:1998bk, Kidonakis:1998nf}\footnote{For simplicity, the argument dependence  of the soft anomalous dimension on the mass scales is suppressed  in Eq.~(\ref{eq:soft:evolfact}).} 
\begin{eqnarray}
\label{eq:soft:evolfact}
\mathbf{\bar{U}}_{ij\to klB}\left(N, Q^2,\{m^2\}, \mur^2\right)  &= & \bar{\mathrm{P}}\exp\left[\int_{\mur}^{Q/\bar{N}}\frac{dq}{q}\mathbf{\Gamma}_{ij\to klB}^{\dagger}\left(\alpha_{\mathrm{s}}\left(q^{2}\right)\right)\right], \\
\mathbf{U}_{ij\to klB}\left(N, Q^2, \{m^2\},\mur^2\right)  &= & \mathrm{P}\exp\left[\int_{\mur}^{Q/\bar{N}}\frac{dq}{q}\mathbf{\Gamma}_{ij\to klB}\left(\alpha_{\mathrm{s}}\left(q^{2}\right)\right)\right], \nonumber
\end{eqnarray}
where $\mathrm{P}$ and  $\bar{\mathrm{P}}$ denote the path- and reverse path-ordering in the variable $q$, respectively. The soft anomalous dimension $\mathbf{\Gamma}_{ij\to klB}$ is a perturbative function in $\als$:
\begin{eqnarray}
\mathbf{\Gamma}_{ij\to klB}\left(\als\right)= \left[ \left(\frac{\als}{\pi}\right) \mathbf{\Gamma}^{(1)}_{ij\to klB} +\left(\frac{\als}{\pi}\right)^2 \mathbf{\Gamma}^{(2)}_{ij\to klB}+\ldots   \right].
\end{eqnarray}
In order to perform resummation at NLL accuracy we need to know $\mathbf{\Gamma}^{(1)}_{ij\to klB}$, whereas NNLL accuracy requires including $\mathbf{\Gamma}^{(2)}_{ij\to klB}$. The one-loop soft anomalous dimension for the process $ij\to klB$ with $k,l$ being heavy quarks can be found e.g.\ in~\cite{Kulesza:2015vda}. The two-loop contributions to the soft anomalous dimension were calculated in~\cite{Ferroglia:2009ep,Ferroglia:2009ii}\footnote{Note that while using the radiative factors as given in~\cite{Catani:1996yz,Catani:2003zt}, we need to subtract the collinear soft radiation already included in $\Delta^i$, $\Delta^j$ from the eikonal cross section used to calculate the soft function.}. In the triple-invariant mass (TIM) kinematics, the soft anomalous dimension matrix in general contains off-diagonal terms, thus complicating the evaluation of the resummed cross section. At NNLL additional difficulty arises because of non-commutativity of $\mathbf{\Gamma}^{(1)}_{ij\to klB}$ and $\mathbf{\Gamma}^{(2)}_{ij\to klB}$ matrices.  

We make use of the method of~\cite{Kidonakis:1998nf} in order to diagonalize the one-loop soft anomalous dimension matrix. Denoting the diagonalization matrix for $\mathbf{\Gamma}^{(1)}_{ij\to klB}$ by $\mathbf{R}$ we have
\begin{equation}
\mathbf{\Gamma}^{(1)}_{R}  =  \mathbf{R}^{-1}\mathbf{\Gamma}^{(1)}_{ij\to klB} \mathbf{R},
\end{equation}
where the diagonalized matrix is given by eigenvalues $\lambda^{(1)}_I$ of $\mathbf{\Gamma}^{(1)}_{ij\to klB}$
$$
\mathbf{\Gamma}^{(1)}_{R,IJ}=\lambda^{(1)}_I \delta_{IJ},
$$ 
and can be also written as  $\mathbf{\Gamma}_R^{(1)}=\left[  \overrightarrow{\lambda}^{\left(1\right)} \right]_D$ with  $\overrightarrow{\lambda}^{\left(1\right)}=\left\{ \lambda^{(1)}_1,\ldots, \lambda^{(1)}_D \right\}$. 
The other matrices are transformed as:
\begin{eqnarray}
\mathbf{\Gamma}^{(2)}_{R} & = & \mathbf{R}^{-1}\, \mathbf{\Gamma}^{(2)}_{ij\to klB}\, \mathbf{R},\nonumber \\
\mathbf{H}_R & = & \mathbf{R}^{-1}\, \mathbf{H}_{ij\to klB} \, \left(\mathbf{R}^{-1}\right)^{\dagger},\\
\mathbf{\tilde S}_{R}& = & \mathbf{R}^{\dagger}\, \mathbf{\tilde S}_{ij\to klB}\, \mathbf{R}.\nonumber 
\end{eqnarray}

At NLL accuracy, by changing the colour basis to the one in which $\mathbf{\Gamma}^{(1)}_{ij\to klB}$ is diagonal,  the path ordered exponentials in Eq.~(\ref{eq:soft:evol}) reduce to sum over simple exponentials. At NNLL accuracy, to recast the path ordered exponential of the soft anomalous dimension matrix in a form containing simple exponential functions, we make use of a technique detailed in e.g.~\cite{Buras:1979yt,Ahrens:2010zv} resulting in
\begin{equation}
\mathbf{U}_R(N, Q^2,\{m^2\}, \mur^2)=\left(\mathbf{1}+\frac{\alpha_{\mathrm{s}}\left(Q^2/\bar{N}^2\right)}{\pi}\mathbf{K}\right)\left[\left(\frac{\alpha_{\mathrm{s}}\left(\mur^2\right)}{\alpha_{\mathrm{s}}\left(Q^2/\bar{N}^2\right)}\right)^{\frac{\overrightarrow{\lambda}^{\left(1\right)}}{2\pi b_{0}}}\right]_{D}\left(\mathbf{1}-\frac{\alpha_{\mathrm{s}}\left(\mur^2\right)}{\pi}\mathbf{K}\right),
\end{equation}
with the subscript $D$ indicating a diagonal matrix. The matrix $\mathbf{K}$ is given by
\begin{equation}
K_{IJ}=\delta_{IJ}{\lambda}^{\left(1\right)}_{I}\frac{b_1}{2b_0^2}-\frac{\left(\mathbf{\Gamma}^{(2)}_R\right)_{IJ}}{2\pi b_0+\lambda^{\left(1\right)}_{I}-\lambda^{\left(1\right)}_{J}}\, ,
\end{equation}
where $b_0$ and $b_1$ are the first two coefficients of expansion $\beta_{\mathrm{QCD}}$ in $\als$:
\begin{eqnarray}
b_0 &=& \frac{11 \CA-4 \nf \TR}{12\pi},\\
b_1 &=& \frac{17 \CA^2-\nf \TR\left(10\CA+6\CF\right)}{24\pi^2}\,.
\end{eqnarray}
In our calculation we set $\nf=5$.

In the diagonal basis of the one-loop soft anomalous dimension, up to NNLL accuracy Eq.~(\ref{eq:res:fact_master})  can be written as
\begin{eqnarray}
\label{eq:res:fact_diag}
\frac{d\tilde \sigh^{{\rm (NNLL)}}_{ij\tosv kl B}}{dQ^2}&&\hspace{-0.9cm}(N, Q^2,\{m^2\}, \mur^2) = {\mathrm{Tr}}\left[ \mathbf{H}_R (Q^2, \{m^2\},\muf^2, \mur^2)\mathbf{\bar{U}}_R(N+1, Q^2,\{m^2\}, Q^2 ) \, \right. \nn
\\ 
&\times& \left. \mathbf{\tilde S_R} (N+1, Q^2, \{m^2\})\, \mathbf{{U}}_R(N+1, Q^2,\{m^2\}, Q^2 )  \right] \nonumber \\
&\times&\,\Delta^i(N+1, Q^2,\muf^2, \mur^2 ) \Delta^j(N+1, Q^2,\muf^2, \mur^2 ). 
\end{eqnarray}

In the above equation, the $\mathbf{H}_R$ and  $\mathbf{\tilde S}_R$ are hard and soft function matrices projected onto $R$ colour basis. They are calculated at the NLO accuracy, i.e. including the ${\cal O} (\als)$ terms in Eqs.~(\ref{eq:hard:nlo}) and (\ref{eq:soft:nlo}). The LO hard matrix is derived from the Born cross section. The NLO hard matrix contains non-logarithmic contributions which are independent of $N$. They consist of virtual loop contributions, real terms of collinear origin and the contributions from the evolution matrices $\mathbf{U_R}$  and $\mathbf{\bar{U}_R}$, corresponding to evolution between $\mur$ and $Q$, expanded up to ${\cal O(\als)}$. The colour-decomposed virtual corrections are extracted from the calculations of the NLO cross section in the {\tt PowHel} framework~\cite{Garzelli:2011vp}.  Aside from evolution terms, the remaining terms in  $\mathbf{H}_R^{(1)}$ are obtained from  the infrared-limit of the real corrections~\cite{Catani:2002hc} using the method initially proposed in~\cite{Beenakker:2011sf,Beenakker:2013mva}. Additionally, we recalculate the one-loop soft function $\mathbf{\tilde S}^{(1)}$~\cite{Ahrens:2010zv, Li:2014ula}. The dependence on $N$ in the soft function $\mathbf{\tilde S}_R$ enters only through the argument of $\als$ in Eq. (\ref{eq:soft:nlo}).


Substituting the expression for the running coupling, we obtain up to NNLL accuracy for the soft matrix evolution factors in Eq.~(\ref{eq:res:fact_diag})
\begin{eqnarray}
\label{eq:UR}
\mathbf{U}_R(N,Q^2,\{m^2\},Q^2 )&=&\left(\mathbf{1}+\frac{\alpha_{\mathrm{s}}(\mur^2)}{\pi(1-2\lambda)}\mathbf{K}\right)
\left[e^{\,g_s(N)\overrightarrow{\lambda}^{(1)}} \right]_{D}
\left(\mathbf{1}-\frac{\alpha_{\mathrm{s}}(\mur^2)}{\pi}\, \mathbf{K}\right),\\
 \mathbf{\bar{U}}_R(N, Q^2,\{m^2\},Q^2 )&=&\left(\mathbf{1}-\frac{\alpha_{\mathrm{s}}(\mur^2)}{\pi}\, \mathbf{K}^{\dagger}\right) 
\left[e^{\,g_s(N)\left(\overrightarrow{\lambda}^{(1)}\right)^*} \right]_{D} \left(\mathbf{1}+\frac{\alpha_{\mathrm{s}}(\mur^2)}{\pi(1-2\lambda)}\mathbf{K}^{\dagger}\right),\nonumber \\
\label{eq:URb}
\end{eqnarray}
where:
\begin{eqnarray}
g_s (N)= \frac{1}{2 \pi b_0} \left\{  \log(1-2\lambda) + \als(\mur^2) \left[   \frac{b_1}{b_0} \frac{ \log(1-2\lambda)}{ 1-2\lambda} - 2 \gamma_{\rm E} b_0  \frac{2\lambda}{1-2\lambda} \right. \right. \nonumber \\
\left. \left.  
+ \, b_0 \log\left( \frac{Q^2}{\mur^2} \right) \frac{2\lambda}{1-2\lambda} \right] \right\},
\label{eq:gsoft}
\end{eqnarray}
and 
\begin{equation}
\lambda = \als(\mur^2) b_0 \log N.
\end{equation}
The $\mathbf{U}_R$ and $\mathbf{\bar{U}}_R$ factors in Eqs.~(\ref{eq:UR}),  (\ref{eq:URb}) correspond to evolution from $Q/\bar{N}$ up to Q and depend on $\mur$ only through the argument $\als$. The $N$-independent evolution from $\mur$ to $Q$  is incorporated into the hard function, as noted earlier.

The other factors contributing to the resummation of logarithms, i.e.\ the radiative factors for the incoming partons, $\Delta^i$ and  $\Delta^j$ are widely known. The results at NLL accuracy  can be found for example in~\cite{Catani:1996yz,Bonciani:1998vc} and at the NNLL level in ~\cite{Catani:2003zt}.

As already noted, at NLL accuracy, by changing the colour basis to $R$-basis,  the path ordered exponentials in Eq.~(\ref{eq:soft:evol}) reduce to simple exponentials. Equivalently, the NLL accuracy can be obtained by neglecting terms suppressed by a factor of $\alpha_{\mathrm{s}}$ in Eqs.~(\ref{eq:UR}),~(\ref{eq:URb})~and~(\ref{eq:gsoft}). This results in the soft matrix evolution factors turning into exponential functions for the eigenvalues of the soft anomalous dimension matrix. At NLL, it is also enough to only know the LO contributions to the hard and soft function, which results in the following expression for the resummed cross section in the Mellin space
\begin{eqnarray}
\frac{d \tilde\sigh^{{\rm (NLL)}}_{ij\tosv kl B}}{dQ^2}&&\hspace{-0.9cm}(N,Q^2,\{m^2\},\muf^2, \mur^2) =  \mathbf{H}^{\mathrm{(0)}}_{R,IJ}(Q^2, \{m^2\}) \, \mathbf{\tilde S}^{\mathrm{(0)}}_{R,JI}\nn\\  &\times&
\,\Delta^i(N+1, Q^2,\muf^2, \mur^2) \Delta^j(N+1, Q^2,\muf^2, \mur^2) \nonumber \\
&\times& \, \exp\left[\frac{\log(1-2\lambda)}{2 \pi b_0}
\left(\left( \lambda^{(1)} _{J}\right)^{*}+\lambda^{(1)} _{I}\right)\right] ,
\label{eq:res:fact_diag_NLL}
\end{eqnarray}
where the color indices $I$ and $J$ are implicitly summed over. The trace of the product of two matrices  $\mathbf{H_R}^{\mathrm{(0)}}$ and $\mathbf{\tilde S}^{\mathrm{(0)}}_{R}$ returns the LO cross section.The incoming parton radiative factors $\Delta_i$ are now considered only at NLL accuracy.

In order to improve the accuracy of the numerical approximation provided by resummation, it is customary to include terms up to $\cal{O}(\als)$ in the expansion of the hard and soft function  leading to 
\begin{eqnarray}
\frac{d \tilde\sigh^{{\rm (NLL\ w\ {\cal C})}}_{ij\tosv kl B}}{dQ^2}&&\hspace{-0.9cm}(N,Q^2,\{m^2\},\muf^2, \mur^2) =  \mathbf{H}_{R,IJ}(Q^2, \{m^2\},\muf^2, \mur^2) \, \mathbf{\tilde S}_{R,JI}(Q^2,\{m^2\})\nn \\
  &\times&
 \Delta^i(N+1, Q^2,\muf^2, \mur^2) \Delta^j(N+1, Q^2,\muf^2, \mur^2) \nonumber \\
&\times& \, \exp\left[\frac{\log(1-2\lambda)}{2 \pi b_0}
\left(\left( \lambda^{(1)} _{J}\right)^{*}+\lambda^{(1)} _{I}\right)\right] .
\label{eq:res:fact_diag_NLLwC}
\end{eqnarray}
where
\beq
\mathbf{H}_{R}\, \mathbf{\tilde S}_{R} = \mathbf{H}^{\mathrm{(0)}}_{R} \mathbf{\tilde S}^{\mathrm{(0)}}_{R} + \frac{\als}{\pi}\left[\mathbf{H}^{\mathrm{(1)}}_{R} \mathbf{\tilde S}^{\mathrm{(0)}}_{R}+ \mathbf{H}^{\mathrm{(0)}}_{R} \mathbf{\tilde S}^{\mathrm{(1)}}_{R} \right] . \nn
\eeq
We will refer to this result as "NLL  w ${\cal C}$", since the $N$-independent $\cal{O}(\als)$ terms in the hard and soft function are often collected together in one function, known as the hard matching coefficient, ${\cal C}$. Although we choose to treat these terms as in Eq.~(\ref{eq:res:fact_diag_NLLwC}), we keep the name "w ${\cal C}$" ("w" standing for "with") as a useful shorthand.

The resummation-improved cross sections for the $pp \to \ttbh$ process are
obtained through matching the  resummed expressions with 
the full NLO cross sections
\bear
\label{hires}
 \frac{d\sigma^{\rm (matched)}_{h_1 h_2 \tosv klB}}{dQ^2}(Q^2,\{m^2\},\muf^2, \mur^2) &=& 
\frac{d\sigma^{\rm (NLO)}_{h_1 h_2 \tosv kl B}}{dQ^2}(Q^2,\{m^2\},\muf^2, \mur^2) \\ \nn
&+&   \frac{d \sigma^{\rm
  (res-exp)}_
{h_1 h_2 \tosv kl B}}{dQ^2}(Q^2,\{m^2\},\muf^2, \mur^2) 
\eear
 with
\bear
\label{invmel}
&& \frac{d \sigma^{\rm
  (res-exp)}_{h_1 h_2 \tosv kl B}}{dQ^2} (Q^2,\{m^2\},\muf^2, \mur^2) \! =   \sum_{i,j}\,
\int_{\sf C}\,\frac{dN}{2\pi
  i} \; \rho^{-N} f^{(N+1)} _{i/h{_1}} (\muf^2) \, f^{(N+1)} _{j/h_{2}} (\muf^2) \nn \\ 
&& \! \times\! \left[ 
\frac{d \tilde\sigh^{\rm (res)}_{ij\tosv kl B}}{dQ^2} (N,Q^2,\{m^2\},\muf^2, \mur^2)
-  \frac{d \tilde\sigh^{\rm (res)}_{ij\tosv kl B}}{dQ^2} (N,Q^2,\{m^2\},\muf^2, \mur^2)
{ \left. \right|}_{\scriptscriptstyle({\rm NLO})}\, \! \right], 
\eear
where "matched" can stand for "NLO+NNLL", "NLO+NLL" or "NLO+NLL w ${\cal C}$" and "res" for "NNLL", "NLL" or "NLL w ${\cal C}$", correspondingly.
The inclusive cross section is obtained by integrating the invariant mass distribution given in Eq.~(\ref{eq:res:fact_diag}) over $Q^2$ and  $ \sigh^{\rm
  (res)}_{ij\tosv kl B}(N,\muf^2, \mur^2) \left. \right|_{\scriptscriptstyle({\rm NLO})}$ represents its perturbative expansion truncated at NLO.
The moments of the parton 
distribution functions (pdf) $f_{i/h}(x, \muf^2)$ are 
defined in the standard way 
$$
f^{(N)}_{i/h} (\muf^2) \equiv \int_0^1 dx \, x^{N-1} f_{i/h}(x, \muf^2). 
$$
The inverse Mellin transform (\ref{invmel}) is evaluated numerically using 
a contour ${\sf C}$ in the complex-$N$ space according to the ``Minimal Prescription'' 
method developed in Ref.~\cite{Catani:1996yz}.

\section{Numerical results for the $pp \to \ttbh$ process at NLO+NNLL accuracy}
\label{s:results}

In this section we present and discuss our state-of-the-art NLO+NNLL predictions for the $\ttbh$ production process at the LHC for two collision energies $\sqrt S=13$ TeV and $\sqrt S=14$ TeV. The results for the total cross section which we present below are obtained by integrating out the invariant mass distribution over invariant mass $Q$. The distribution in $Q$ undergoes resummation of soft gluon corrections in the threshold limit $\shat \to Q^2$, i.e.\ in the invariant mass kinematics. This approach is different from directly resumming corrections to the total cross section in the absolute threshold limit $\shat \to M^2$, which we performed in~\cite{Kulesza:2015vda}. Numerical results involving resummation  are obtained using two independently developed in-house computer codes. Apart from NLL and NNLL predictions matched to NLO according to Eq.~(\ref{hires}), we also show the NLL predictions supplemented with the $\cal O(\als)$ non-logarithmic contributions ("NLL w ${\cal C}$"), also matched to NLO.

In the phenomenological analysis we use $m_t=173$ GeV and $m_H=125$ GeV. The NLO cross section is calculated using the aMC@NLO code~\cite{Alwall:2014hca}. We perform the current analysis employing PDF4LHC15{\_}100 sets~\cite{Butterworth:2015oua,Dulat:2015mca,Harland-Lang:2014zoa,Ball:2014uwa,Gao:2013bia,Carrazza:2015aoa} and use the corresponding values of $\als$. In particular, for the NLO+NLL predictions we use the NLO sets, whereas the NLO+NNLL predictions are calculated with NNLO sets. For the sake of comparison with Broggio et al.~\cite{Broggio:2016lfj}, we adopt the same choice of pdfs, i.e. MMHT2014~\cite{Harland-Lang:2014zoa}.

We present most of our analysis for two choices of the central values of the renormalization and factorization scales:  $\mu_0=\mufo=\muro=Q$ and $\mu_0=\mufo=\muro=M/2$. The former choice is motivated by invariant mass $Q$ being the natural scale for the invariant mass kinematics used in resummation. The latter choice of the scale is often made in the NLO calculations of the total cross section reported in the literature, see e.g.~\cite{Dittmaier:2011ti}. By studying results for these two relatively distant scales, we aim to cover a span of scale choices relevant in the problem.  The theoretical error due to scale variation is calculated using the so called 7-point method, where the minimum and maximum values obtained with $(\muf/\mu_{0}, \mur/\mu_{0}) = (0.5,0.5), (0.5,1), (1,0.5), (1,1), (1,2), (2,1), (2,2)$ are considered. For reasons of technical simplicity, the pdf error is calculated for the NLO predictions, however we expect that adding the soft gluon corrections only minimally influences the value of the pdf error.

As discussed in the previous section for the evaluation of the first-order hard function matrix $\mathbf {H}^{(1)}_{IJ}$ we need to know one-loop virtual corrections to the process, decomposed into various colour transitions $IJ$. We extract them numerically using the publicly available {\tt PowHel} implementation of the $\ttbh$ process~\cite{Garzelli:2011vp}. In particular, we use analytical relations to translate between virtual corrections split into various colour configurations in the colour flow basis used in~\cite{Garzelli:2011vp} and our default singlet-octet(s) bases. We cross check the consistency of results obtained in this way by comparing the colour-summed one-loop virtual contributions to ${\mathrm{Tr}}\left[\mathbf{H^{(1)}\tilde S^{(0)}}\right]$ with the full one-loop virtual correction given by the {\tt PowHel} package~\cite{Garzelli:2011vp}, as well as the {\tt POWHEG} implementation of the $\ttbh$ process~\cite{Hartanto:2015uka} and the standalone MadLoop implementation in {\tt aMC@NLO}~\cite{Hirschi:2011pa}. 

\begin{figure}[h]
\centering
\includegraphics[width=0.45\textwidth]{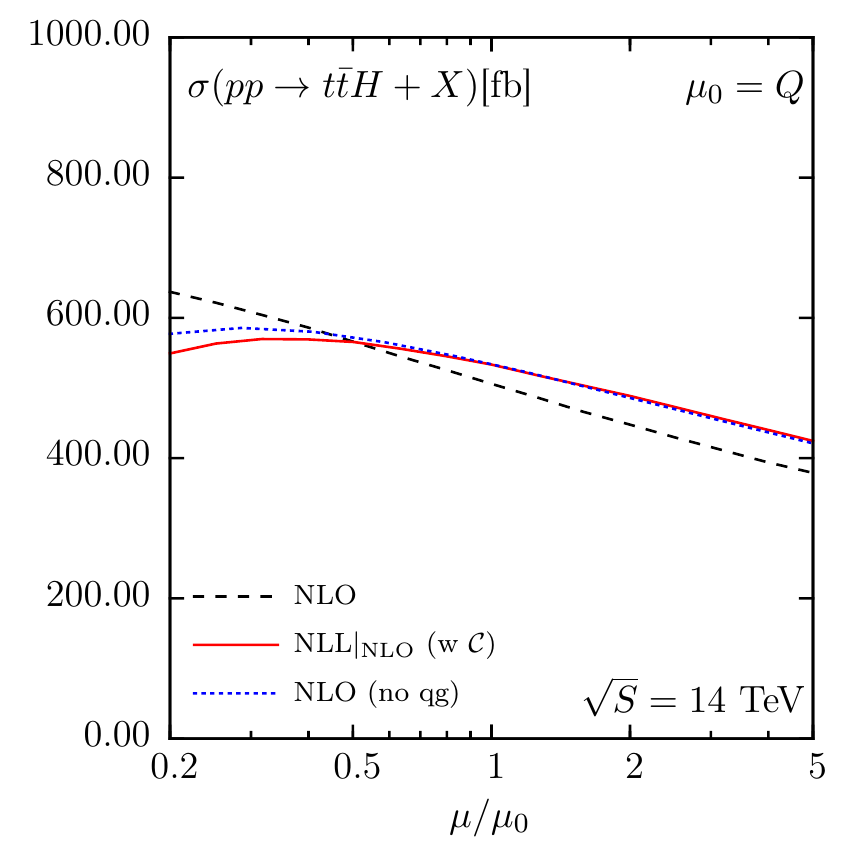}
\includegraphics[width=0.45\textwidth]{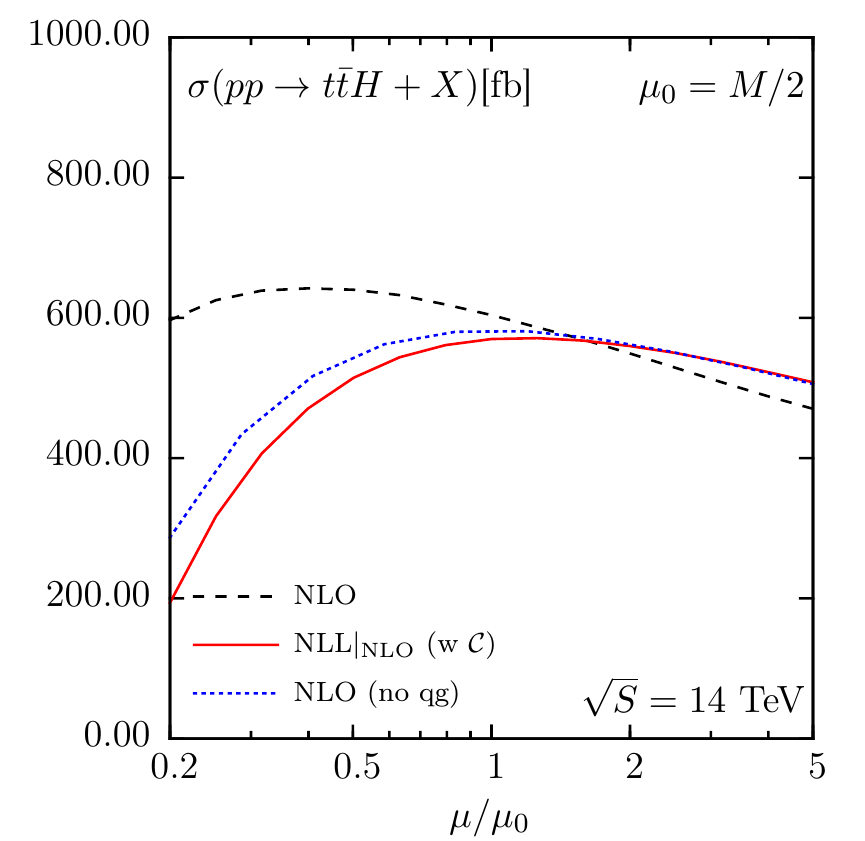}
\caption{Comparison between the expansion of the resummed expression Eq.~(\ref{eq:res:fact_diag}) up to NLO accuracy in $\alpha_{\mathrm{s}}$, the full NLO result and the NLO result without the $qg$ channel contribution as a function of the scale $\mu=\muf=\mur$.} 
\label{f:NLL_expansion_vs_NLO}
\end{figure}

We begin the discussion of numerical results by analyzing how well the full NLO result for the total cross section is approximated by the expansion of the resummed cross section up to the same accuracy in $\als$ as in NLO. It was first pointed out in~\cite{Li:2014ula} in the context of the $t \bar t W$ production and then later in~\cite{Broggio:2015lya} for the $t \bar t H$ process that the $qg$ production channel carries a relatively large numerical significance, especially in relation to the scale uncertainty. This is due to the fact that a non-zero contribution from the $qg$ channel appears first at NLO, i.e.\ it is subleading w.r.t.\ contributions from the $q \bar q$ and $gg$ channels. Correspondingly, no resummation is performed for this channel and it enters the matched resummation-improved formula~Eq.~(\ref{hires}) only through a fixed order contribution at NLO. It is then clear that in order to estimate how much of the NLO result is constituted by the terms accounted for in the resummed expression, Eq.~(\ref{eq:res:fact_diag}), its expansion  should not be directly compared with full NLO but with NLO cross section without a contribution from the $qg$ channel. We obtain the latter result from the {\tt PowHel} package~\cite{Garzelli:2011vp}. Its comparison with the expansion of the resummed expression in Eq.~(\ref{eq:res:fact_diag}) up to NLO accuracy in $\als$ as a function of the scale $\mu=\muf=\mur$ is shown in Figure~\ref{f:NLL_expansion_vs_NLO} for $\sqrt S=14$ TeV and two choices of the central scale $\mu_0=Q$ and $\mu_0=M/2$. While in both cases the expansion of the resummed cross section differs significantly from the full NLO, the NLO result with the $qg$ channel contribution subtracted is much better approximated by the expansion, especially for the dynamical scale choice $\mu=Q$ and for the fixed scale choice $\mu \geq M/2$, for the physically motivated scale choices. Such good agreement lets us conclude that the NNLL resummed formula will indeed take into account a prevailing part of the higher-order contributions from the $q \bar q$ and $gg $ channels to all orders in $\als$.\footnote{Although the expansion and the NLO results w/o the $qg$ channel contribution agree very well at this level of accuracy in $\als$, since we do not know the second-order hard-matching coefficients we cannot expect an equally good approximation of the NNLO result by the expansion of the NNLL formula.} 

\begin{figure}[h]
\centering
\includegraphics[width=0.45\textwidth]{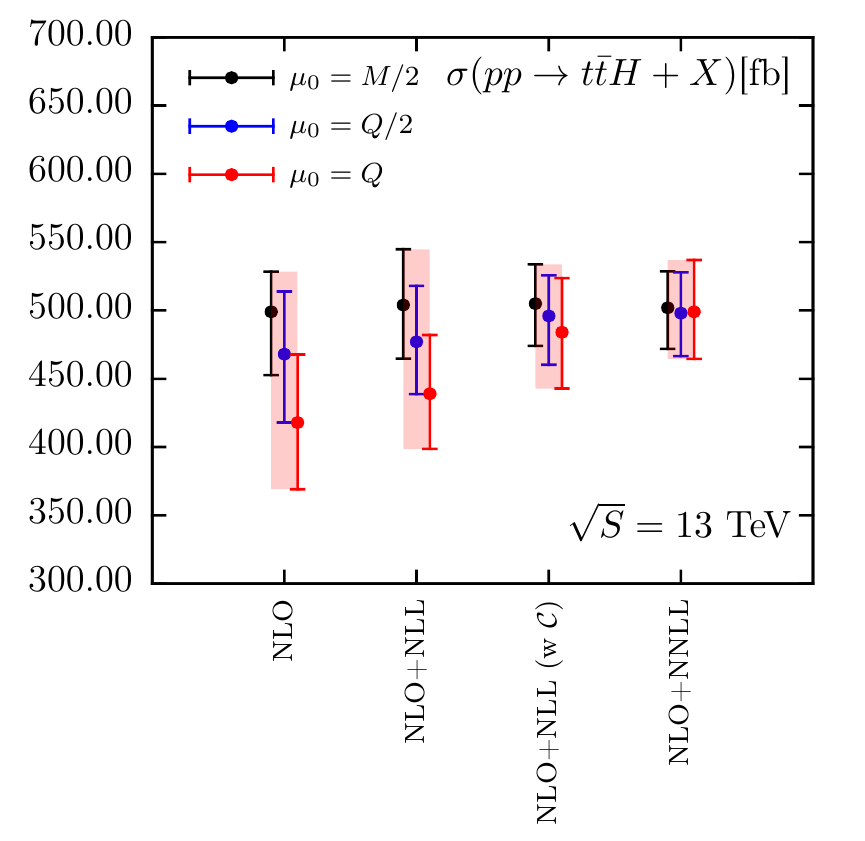}
\includegraphics[width=0.45\textwidth]{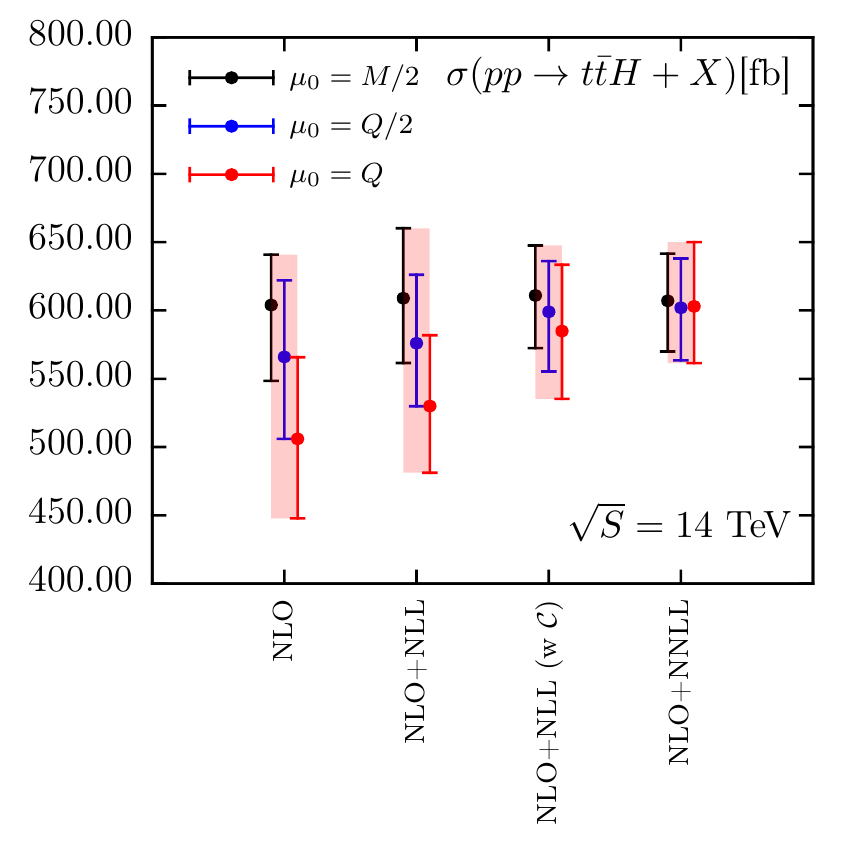}
\caption{Graphical illustration of results presented in Table~\ref{t:totalxsec}.} 
\label{f:totalxsec}
\end{figure}

Our numerical predictions for the total cross sections at $\sqrt S=13$ TeV and $\sqrt S=14$ TeV are shown in Table~\ref{t:totalxsec}. We report results obtained with our default scale choice $\mu_0=Q$ as well as the fixed scale $\mu_0=M/2$. Additionally, we also provide results for the `in-between' choice of $\mu_0=Q/2$. 
While for these choices of central scale the NLO results vary by 20 \%, the variation\footnote{The value of 10\% scale error often quoted in the literature relates to a variation by factors of 0.5 or 2 around $\mu_0=M/2$, while here we consider a much wider range between $M/2$ and $Q$.} reduces as the accuracy of resumation increases. In particular, the NLO+NNLL results show a remarkable stability w.r.t.\ the scale choice. We also observe that the 7-point method scale uncertainty of the results gets reduced with the increasing accuracy. In particular, for all scale choices, the scale uncertainty of NLO+NNLL cross section is reduced compared to the NLO scale uncertainty calculated in the same way. The degree up to which the scale uncertainty is reduced depends on the specific choice of the central scale. For example, for $\mu_0=Q/2$ the theoretical precision of the NLO+NNLL prediction is improved by about 40\% with respect to the NLO result, bringing the scale error calculated with the 7-point method down to less than 6.5\% of the central cross section value. The results shown in Table~\ref{t:totalxsec} are further graphically presented in Fig.~\ref{f:totalxsec}.

\begin{table}
\begin{center}
\begin{tabular}{|c c c c c  c |}
	\hline
	$\sqrt{S}$ {[}TeV{]} & $\mu_0$ & NLO {[}fb{]} & {NLO+NLL}{[}fb{]} & {NLO+NLL with $\cal C$} {[}fb{]} & {NLO+NNLL}{[}fb{]} \tabularnewline
	\hline	 
   13 & $Q$ & $418_{-11.7\%}^{+11.9\%}$ & $439_{-9.2\%}^{+9.8\%}$  & $484_{-8.5\%}^{+8.2\%}$  & $499_{-6.9\%}^{+7.6\%}$   \tabularnewline
	  & $Q/2$ & $468_{-10.7\%}^{+9.8\%}$ & $477_{-8.0\%}^{+8.6\%}$  & $496_{-7.2\%}^{+6.0\%}$  & $498_{-6.3\%}^{+6.0\%}$  \tabularnewline
	 & $M/2$ & $499_{-9.3\%}^{+5.9\%}$ & $504_{-7.8\%}^{+8.1\%}$  & $505_{-6.1\%}^{+5.7\%}$  & $502_{-6.0\%}^{+5.3\%}$  \tabularnewline	
	\hline 
	14  & $Q$ & $506_{-11.5\%}^{+11.8\%}$ & $530_{-9.2\%}^{+9.8\%}$  & $585_{-8.5\%}^{+8.3\%}$ &  $603_{-6.9\%}^{+7.8\%}$  \tabularnewline
	 & $Q/2$ & $566_{-10.6\%}^{+9.9\%}$ & $576_{-8.0\%}^{+8.7\%}$  & $599_{-7.3\%}^{+6.2\%}$  & $602_{-6.4\%}^{+6.0\%}$  \tabularnewline 
	& $M/2$ & $604_{-9.2\%}^{+6.1\%}$ & $609_{-7.8\%}^{+8.4\%}$  & $611_{-6.3\%}^{+6.0\%}$  & $607_{-6.1\%}^{+5.7\%}$  \tabularnewline
	\hline 
\end{tabular}
\end{center}
\caption{Total cross section predictions for $pp \to \ttbh$ at various LHC collision energies and central scale choices. The listed error is the theoretical error due to scale variation calculated using the 7-point method.}
\label{t:totalxsec}
\end{table}

The size of the $K_{\mathrm{NNLL}}$ factor measuring the impact of the higher-order logarithmic corrections,  defined as the ratio of the NLO+NNLL to NLO cross sections, is shown in Table~\ref{t:totalxsecKfactor}. It varies depending on the value of the central scale. The variation is almost entirely driven by the scale dependence of the NLO cross section. For the choice $\mu_0=Q$ the $K_{\mathrm{NNLL}}$-factor can be as high as 1.19.

Given the conspicuous stability of the NLO+NNLL results, see Fig.~\ref{f:totalxsec}, we are encouraged to combine our results obtained for various scale choices. For this purpose we adopt the method proposed by the Higgs Cross Section Working Group~\cite{Dittmaier:2011ti}. In this way, we obtain for the $\ttbh$ cross section at 13 TeV
$$
\sigma_{\rm NLO+NNLL}=500 ^{+7.5 \% + 3.0\%}_{-7.1\% -3.0\%} \ {\rm fb},
$$
and at 14 TeV
$$
\sigma_{\rm NLO+NNLL}=604 ^{+7.6 \% + 2.9\%}_{-7.1\% -2.9\%} \ {\rm fb},
$$
where the first error is the theoretical uncertainty due to scale variation and the second error is the pdf uncertainty.

\begin{table}
\begin{center}
\begin{tabular}{|c c c c|}
	\hline
	$\sqrt{S}$ {[}TeV{]} & $\mu_0$ &  {NLO+NNLL }[fb] & $K_{\rm NNLL}$ factor \tabularnewline
	\hline	 
   13 & $Q$   & $499_{-6.9\% - 2.9\%}^{+7.6\% + 2.9\%}$ & 1.19  \tabularnewline
	  & $Q/2$   & $498_{-6.3\%-3.0\%}^{+6.0\%+3.0\%}$ & 1.06  \tabularnewline
	 & $M/2$  & $502_{-6.0\%-3.1\%}^{+5.3\%+3.1\%}$ & 1.01  \tabularnewline	
	\hline 
	14  & $Q$  &  $603_{-6.9\%-2.8\%}^{+7.8\%+2.8\%}$ & 1.22  \tabularnewline
	 & $Q/2$   & $602_{-6.4\%-2.9\%}^{+6.0\%+2.9\%}$ & 1.06  \tabularnewline 
	& $M/2$   & $607_{-6.1\%-3.0\%}^{+5.7\%+3.0\%}$ & 1.01  \tabularnewline
	\hline 
\end{tabular}
\end{center}
\caption{Total cross section predictions at NLO+NNLL for $pp \to \ttbh$ at various LHC collision energies and central scale choices. The first error is the theoretical error due to scale variation calculated using the 7-point method and the second is the pdf error.}
\label{t:totalxsecKfactor}
\end{table}

\begin{figure}[h]
\centering
\includegraphics[width=0.45\textwidth]{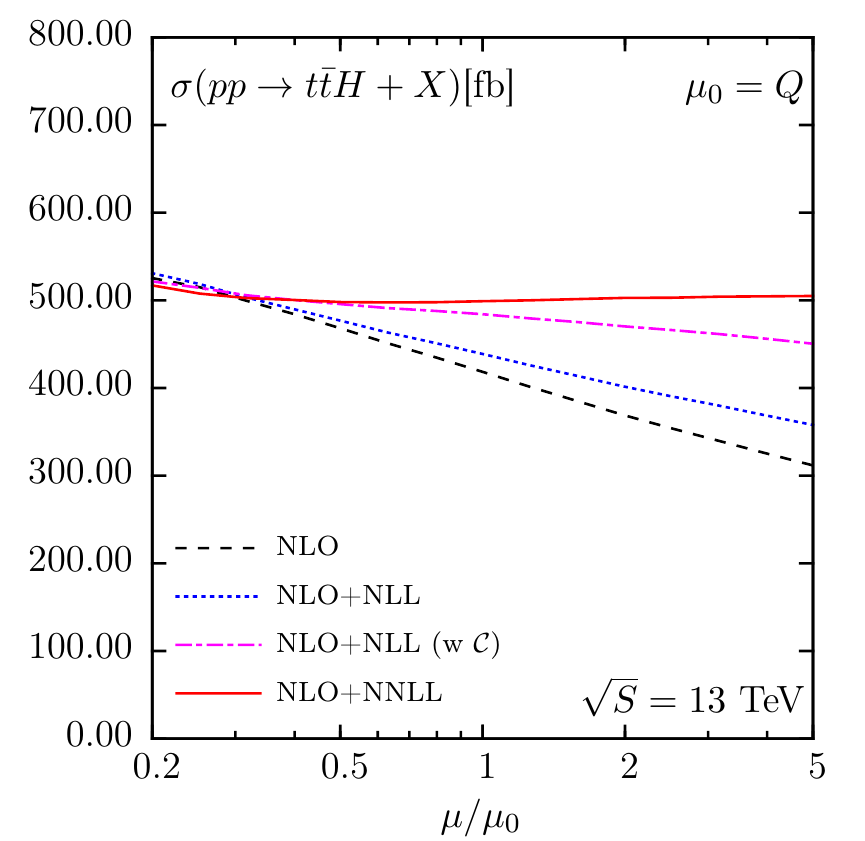}
\includegraphics[width=0.45\textwidth]{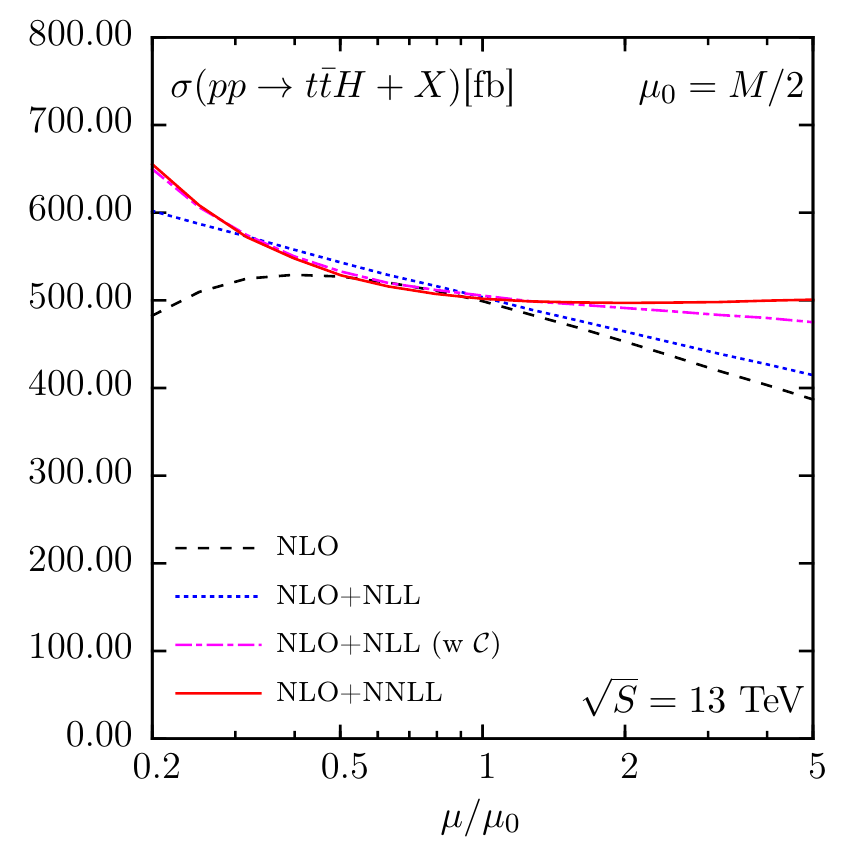}
\caption{Scale dependence of the total cross section for the process $pp\to \ttbh$ at the LHC with $\sqrt S=13$ TeV. Results shown for the choice $\mu=\muf=\mur$ and two central scale values $\mu_0=Q$ (left plot) and $\mu_0=M/2$ (right plot).} 
\label{f:scaledependence13}
\end{figure}

\begin{figure}[h]
\centering
\includegraphics[width=0.45\textwidth]{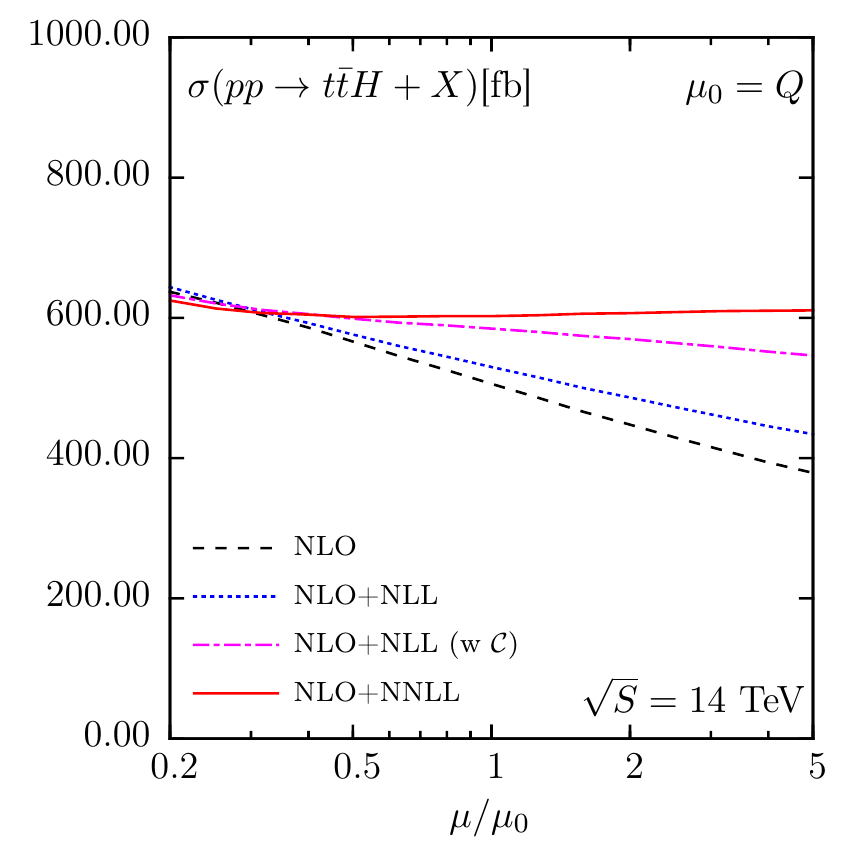}
\includegraphics[width=0.45\textwidth]{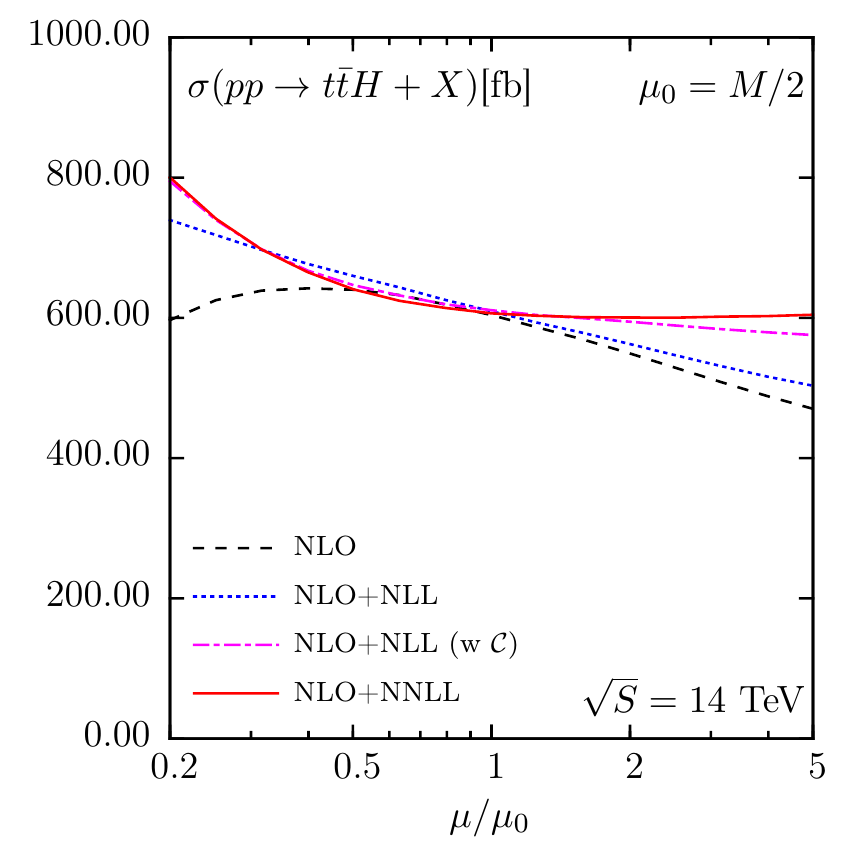}
\caption{Scale dependence of the total cross section for the process $pp\to \ttbh$ at the LHC with $\sqrt S=14$ TeV. Results shown for the choice $\mu=\muf=\mur$ and two central scale values $\mu_0=Q$ (left plot) and $\mu_0=M/2$ (right plot).} 
\label{f:scaledependence14}
\end{figure}

Our findings are further illustrated in the plots in Fig.~\ref{f:scaledependence13} and Fig.~\ref{f:scaledependence14}. We show there the scale dependence of $\ttbh$ total cross sections calculated with the factorization and renormalization scale kept equal, $\mu=\muf=\mur$ for two LHC collision energies $\sqrt S=13$ TeV and $\sqrt S=14$ TeV.  As readily expected, apart from quantitative differences there is no visible disparity between  the qualitative behaviour of results for the two energies. For the central scale choice of $\mu_0=Q$, we observe a steady increase in the stability of the cross section value w.r.t.\ scale variation as the accuracy of resummation improves from NLO+NLL to NLO+NNLL. Our final NLO+NNLL prediction is characterised by a very low scale dependence if $\muf=\mur$ choice is made. Correspondingly, if calculated only along the $\muf=\mur$ direction, the theoretical error on the NLO+NNLL prediction due to scale variation would be  at the level of 1\%, which is a significant reduction from the 10\% variation of the NLO, c.f. Table~\ref{t:totalxsec}.  Results obtained with the scale choice of $\mu_0=M/2$ behave mostly in a similar way. Only in the very low scale regime, $\mu \lesssim 0.2 M$,  the NLO+NNLL cross section shows a stronger scale dependence. For this scale choice, the rise of the matched resummed predictions with the diminishing scale is driven by the fall of the expanded resummed NLL$|_{\rm NLO}$ results, cf.\ Fig.\ref{f:NLL_expansion_vs_NLO}, and a therefore is a consequence of the relatively large scale dependence of NLO contributions stemming from the $qg$ channel. 

\begin{figure}[h]
\centering
\includegraphics[width=0.45\textwidth]{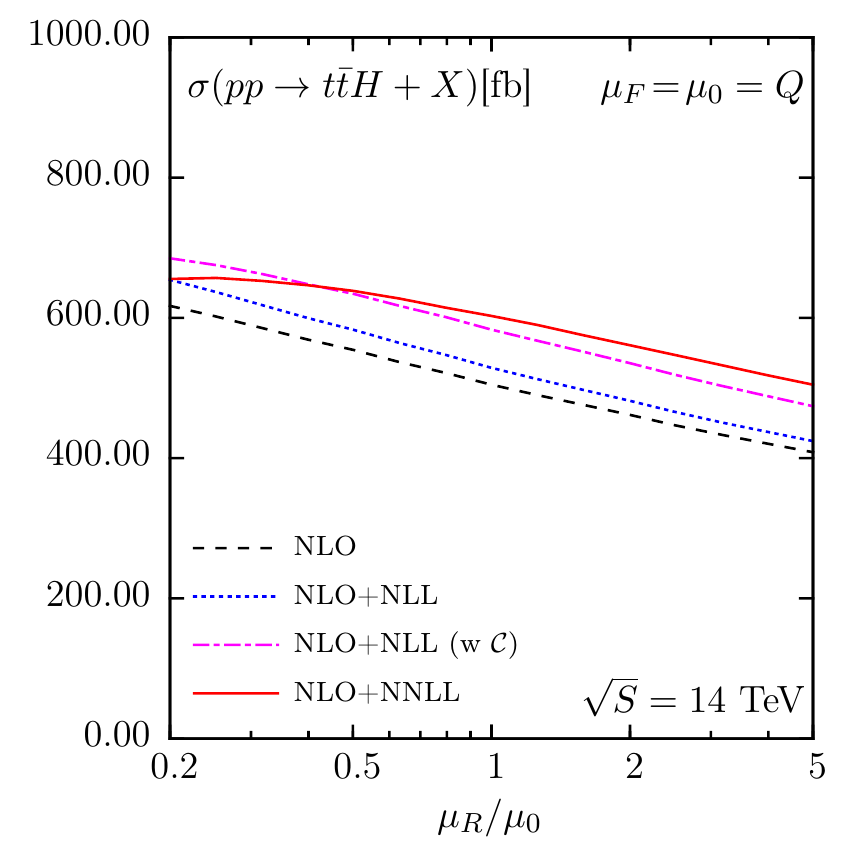}
\includegraphics[width=0.45\textwidth]{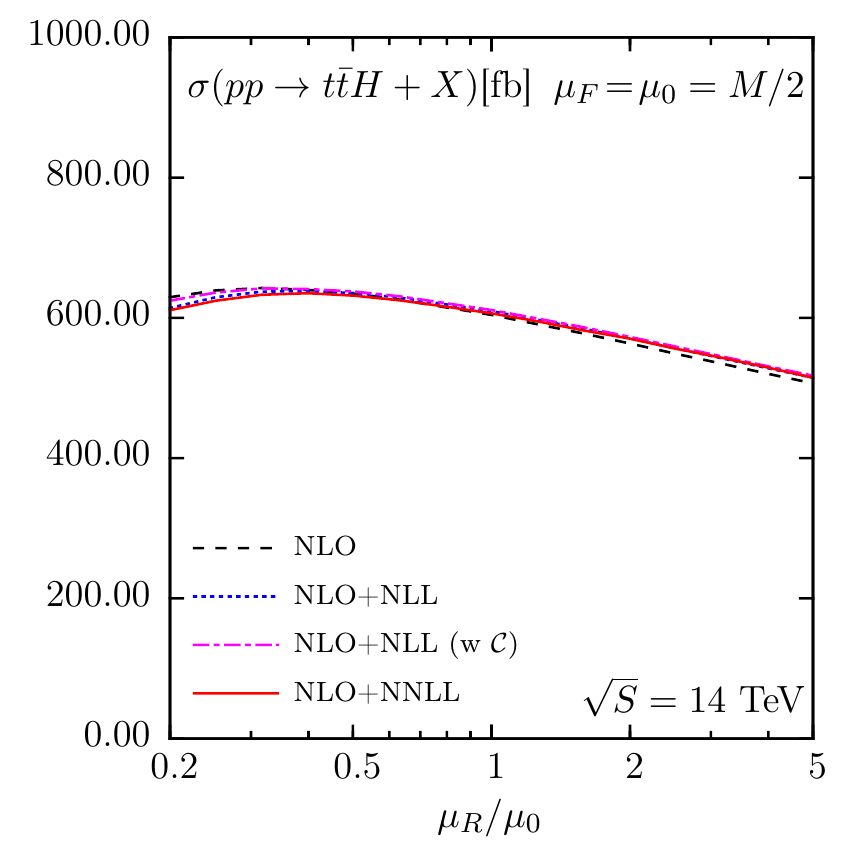}
\caption{Renormalization scale dependence of the total cross section for the process $pp\to \ttbh$ at the LHC with $\sqrt S=14$ TeV and $\muf=\mufo$ kept fixed. Results shown two central scale choces $\mu_0=\mufo=\muro=Q$ (left plot) and $\mu_0=\mufo=\muro=M/2$ (right plot).} 
\label{f:murdep}
\end{figure}

\begin{figure}[h]
\centering
\includegraphics[width=0.45\textwidth]{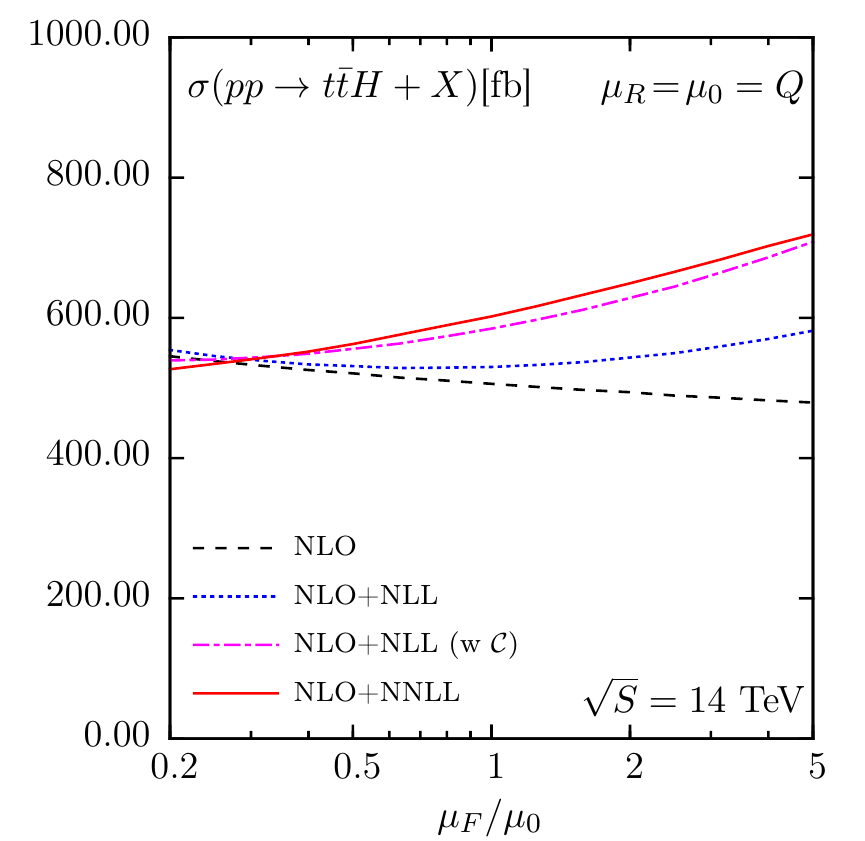}
\includegraphics[width=0.45\textwidth]{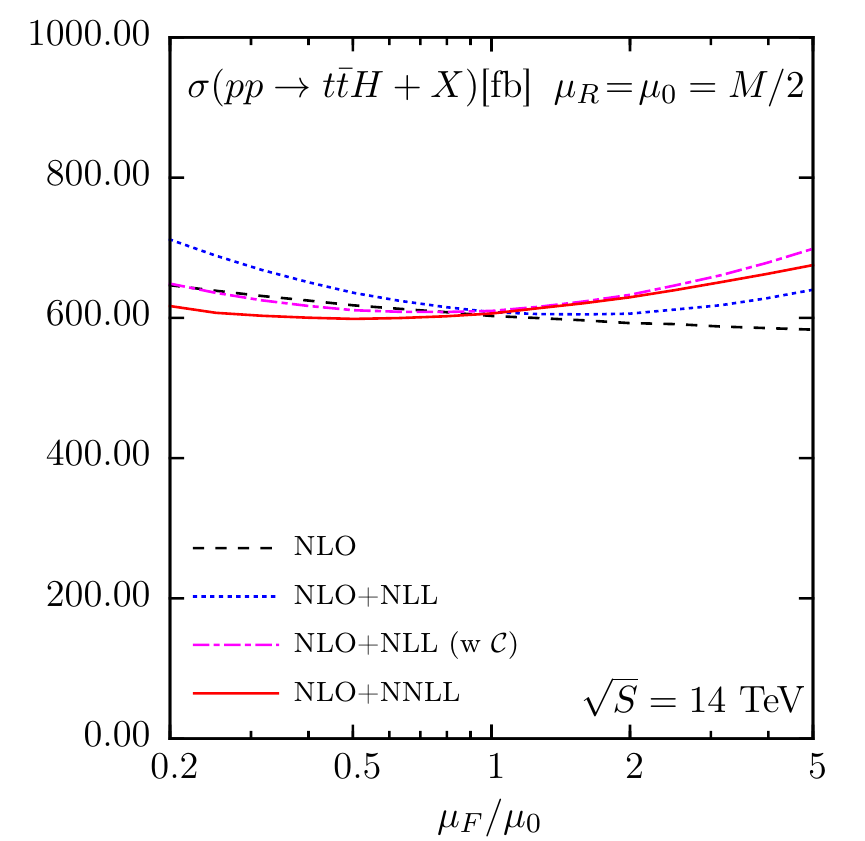}
\caption{Factorization scale dependence of the total cross section for the process $pp\to \ttbh$ at the LHC with $\sqrt S=14$ TeV and $\mur=\mu_{R,0}$ kept fixed. Results shown two central scale choces $\mu_0=\mufo=\muro=Q$ (left plot) and $\mu_0=\mufo=\muro=M/2$ (right plot).} 
\label{f:mufdep}
\end{figure}

We further investigate the dependence on the scale but showing separately the renormalization and factorization scale dependence while keeping the other scale fixed. Fig.~\ref{f:murdep} shows the dependence on $\mur$ and  Fig.~\ref{f:mufdep} on $\muf$ for the $\sqrt S=14$ TeV. We conclude that the weak scale dependence present when the scales are varied simultaneously is a result of the opposite behaviour of the total cross section under $\muf$ and $\mur$ variations. The effect is similar to the cancellations between renormalization and factorization scale dependencies for threshold resummation in the absolute threshold limit which we observed in~\cite{Kulesza:2015vda}. The typical decrease of the cross section with increasing $\mur$ originates from running of $\als$. The behaviour under variation of the factorization scale, on the other hand, is related to the effect of scaling violation of pdfs at probed values of $x$. In this context, it is interesting to observe that the NLO+NLL predictions in Fig.~\ref{f:mufdep} show very little $\muf$ dependence around the central scale, in agreement with expectation of the factorization scale dependence in the resummed exponential and in the pdfs cancelling each other, here up to NLL. The relatively strong dependence on $\muf$ of the NLO+NLL (w $\cal{C}$) predictions  can be then easily understood: the resummed expression will take into account higher-order scale-dependent terms which involve higher-order terms of both logarithmic (in $N$) and non-logarithmic origin. The latter terms do not have their equivalent in the pdf evolution since the pdfs do not carry any process-specific information. Correspondingly, the $\muf$ dependence does not cancel and can lead to strong effects if the non-logarithmic terms are numerically significant.

\begin{figure}[h]
\centering
\includegraphics[width=0.45\textwidth]{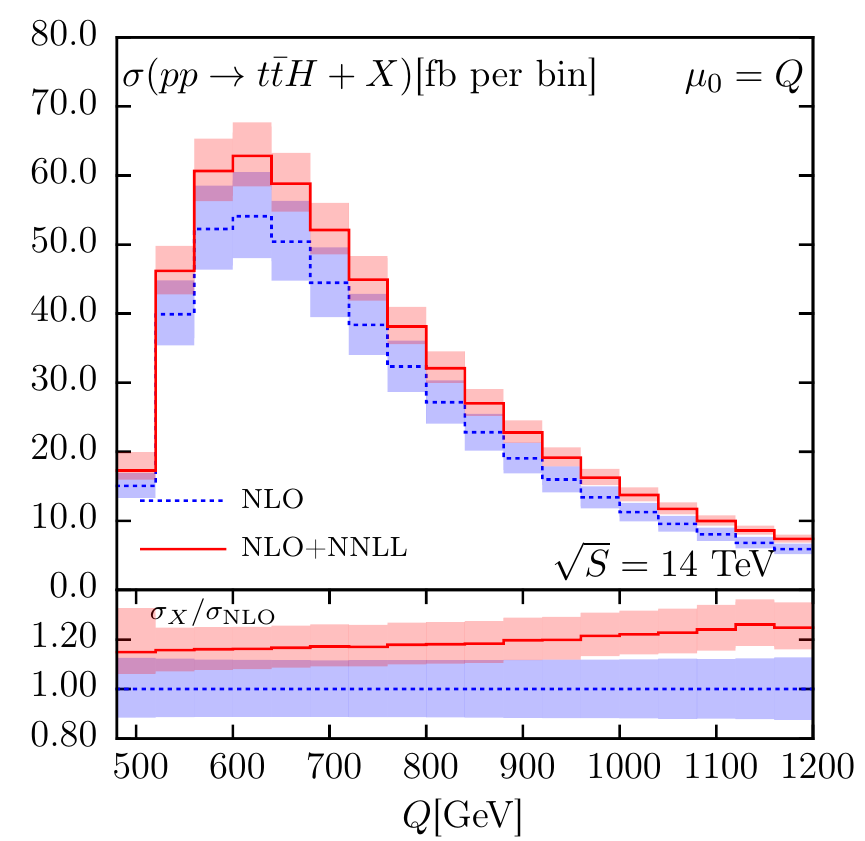}
\includegraphics[width=0.45\textwidth]{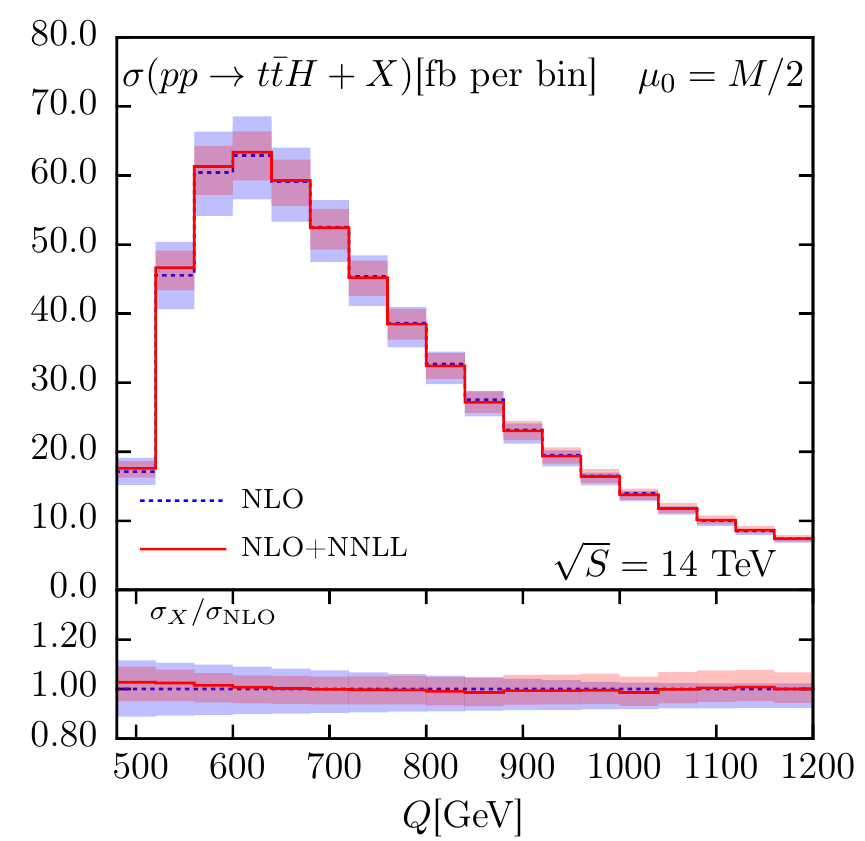}
\caption{Comparison of the NLO+NNLL and NLO invariant mass distributions for the process $pp \to \ttbh$ at the LHC with $\sqrt S=14$ TeV. Results shown for  two central scale choices $\mu_0=Q$ (left plot) and $\mu_0=M/2$ (right plot). Lower panels show the ratio of the distributions w.r.t.\ the NLO predictions.} 
\label{f:Qdiff}
\end{figure}

Given apparent cancellations between $\mur$ and $\muf$ scale dependence, we believe that the 7-point method of estimating the scale error, allowing for an independent variation of $\mur$ only (for $\muf=\mu_0$), is better suited here as an estimate of the theory error than the often used variation of $\mu=\muf=\mur$. Another reason for our preference of this conservative estimate is presence of the hard and soft functions in the resummation formula, Eq.~(\ref{eq:res:fact_diag}), which involve virtual corrections and are known only up to the order $\alpha_{\mathrm{s}}$. Due to suppression of the LO phase space, they provide a relatively significant part of the NLO+NNLL corrections to the total cross sections, cf.\ Table~\ref{t:totalxsec}. It is then justified to suppose that a similar situation might take place also at higher logarithmic orders and that the value of the yet unknown two-loop virtual corrections which feed into the second-order coefficients in  Eq.~(\ref{eq:res:fact_diag}) can have a non-negligible impact on the predictions. With the 7-point method error estimate, we expect that this effect is included within the size of the error.

\begin{figure}[h]
\centering
\includegraphics[width=0.45\textwidth]{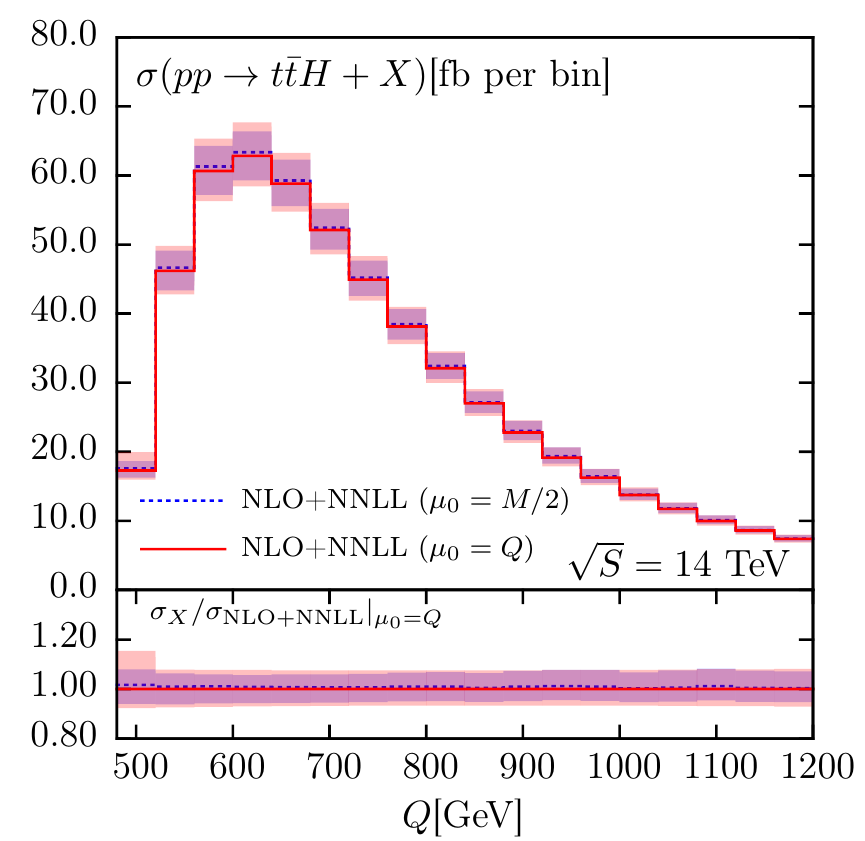}
\caption{Comparison of the NLO+NNLL  invariant mass distributions for the process $pp \to \ttbh$ at the LHC with $\sqrt S=14$ TeV calculated with $\mu_0=Q$ and $\mu_0=M/2$. Lower panel shows the ratio to the NLO+NNLL distribution with $\mu_0=Q$.} 
\label{f:Qdiff_NNLLcomp}
\end{figure}

Our observation of stability of the predictions w.r.t.\ scale variation is also confirmed at the differential level. In Fig.~\ref{f:Qdiff} we show the differential distribution in the invariant mass $Q$ of the $\ttbh$ system produced at  $\sqrt S=14$ TeV. While the NLO distributions calculated with $\mu_0=Q$ and $\mu_0=M/2$ differ visibly, the NLO+NNLL distributions for these scale choices are very close in shape and value. The stability of the NLO+NNLL distribution w.r.t.\ the scale choice  is  demonstrated explicitly in Fig.~\ref{f:Qdiff_NNLLcomp}. Correspondingly, the ratios of the NLO+NNLL to NLO distributions differ. In particular for the choice of $\mu_0=Q$ the NNLL differential K-factor grows with the invariant mass and can be higher than 1.2 at large $Q$. The scale error for the invariant mass distribution is also calculated using the 7-point method. The error bands are slightly narrower for the NLO+NNLL distributions than at NLO.  If the scale errors were calculated by variation of $\mu=\muf=\mur$ by factors of 0.5 and 2, the NLO+NNLL error bands would be considerably narrower.

We complete this part of the discussion by comparing resummed results  obtained using the invariant mass kinematics with those obtained earlier by us in the absolute mass threshold limit~\cite{Kulesza:2015vda}. At 13 (14)~TeV, our most accurate prediction in these kinematics, i.e.\ the NLO+NLL cross section including the first-order hard-matching coefficient, evaluated with PDF4LHC15{\_}100 pdf sets, amounted to $\sigma_{\rm NLO+NLL\,w\,{\cal C}}^{}=530^{+7.8\%}_{-5.5\%} \; (641^{+7.9\%}_{-5.5\%}) $. The absolute mass threshold approach allows only for a fixed scale choice, which is taken to be $\mu_0=\muf=\mur=M/2$.  Comparing this result with our NLO+NLL predictions for the same scale choice in the invariant mass kinematics, cf.\ Table~\ref{t:totalxsec}, we see that the results calculated using the two resummation methods agree within errors.

\begin{figure}[h]
\centering
\includegraphics[width=0.45\textwidth]{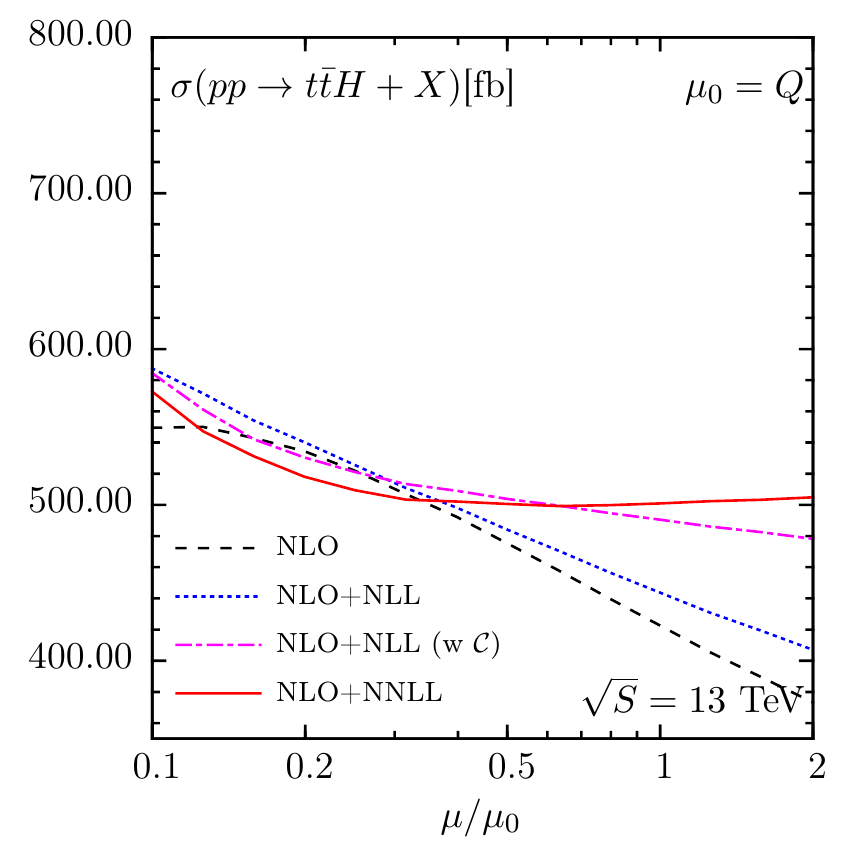}
\caption{Scale dependence of the total cross section for the process $pp\to \ttbh$ at the LHC with $\sqrt S=13$ TeV, calculated using MMHT 2014 pdfs. Results shown for the choice $\mu=\muf=\mur$ and the central scale value $\mu_0=Q$. } 
\label{f:SCET:comp}
\end{figure}

In the remaining part of this section we comment on the relation of our results to the results of Broggio et al. \cite{Broggio:2016lfj}. That work relies on a resummation formula derived in the SCET framework in~ \cite{Broggio:2015lya}, though for the purpose of numerical calculations the Mellin space is adopted. In order to facilitate a comparison with results of \cite{Broggio:2016lfj} we recalculate our results as a function of the scale $\mu=\muf=\mur$ using MMHT2014 pdfs as in \cite{Broggio:2016lfj}. The outcome is presented in Fig.~\ref{f:SCET:comp}, where we show the NLO cross section and the matched resummed cross sections at various accuracy as a function of $\mu=\muf=\mur$ for the range of scales same as in Fig.~1 of \cite{Broggio:2016lfj}. Comparing the two figures, we find a qualitatively similar behaviour of the NLO+NNLL cross sections as a function of the scale. Likewise, we obtain bigger NNLL corrections for the  $\mu_0=\muf=\mur =Q$ scale choice than for $\mu_0=\muf=\mur =Q/2$. However, our NLO+NNLL results appear to be more stable wrt. the scale variation, leading to very little difference between the predictions  for  $\mu_0=Q/2$ and $\mu_0=Q$ (cf.  also Table~\ref{t:totalxsec}). Fig.~\ref{f:SCET:comp} additionally illustrates another feature of our results, namely that for physically relevant values of $\mu_0 \gtrsim 0.3\, Q$ the scale dependence diminishes as the accuracy of the predictions increases, independently on the choice of the central scale $\mu_0$.

However, it has to be noted that the scale choices made to obtain results reported in this paper and \cite{Broggio:2016lfj} are not equivalent. While our resummed expressions depend on $\muf$ and $\mu_R$, the formulas used in \cite{Broggio:2016lfj} contain dependence on the hard and soft scales $\mu_h$ and $\mu_s$, as well as $\muf$. The $\mu_s$ scale in \cite{Broggio:2016lfj} is chosen in such a way as to mimic the scale of soft radiation in the Mellin-space framework, i.e. $\mu_s=Q/\bar N$. Furthermore, for a given $\muf$ the resummed central results of \cite{Broggio:2016lfj} are obtained with a fixed hard scale $\mu_h=Q$, while the exact and approximate NLO results are evaluated keeping all other scales equal to the factorization scale. 
There is one choice of factorization scale for which the scale setting procedure of \cite{Broggio:2016lfj} corresponds to simultaneous variation of $\mu=\muf=\mu_R$, that is $\muf=Q$.  For this choice we obtain $\sigma_{\rm NLO+NNLL}=501.7^{+38.6}_{-34.6}$ fb, to be compared with $514.3^{+42.9}_{-39.5}$ fb reported in \cite{Broggio:2016lfj}, i.e. the central results of the two calculations agree within 2.5\%. (The scale errors given together with the central values are expected to vary due to the different methods used for calculating them.)  At NLO+NLL accuracy we do not find an agreement with \cite{Broggio:2016lfj}. We conclude that the differences in properties of the NLO+NNLL cross sections reported here and in \cite{Broggio:2016lfj}  are likely related  to handling  of scale setting in the two resummation approaches.

\section{Summary}

In this work, we have investigated the impact of the soft gluon emission effects on the total cross section for the process $pp\to t\tb H$ at the LHC. The resummation of soft gluon emission has been performed using the Mellin-moment resummation technique at the NLO+NNLL accuracy in the three particle invariant mass kinematics. We have considered the differential distribution in the invariant mass as well as the total cross section, obtained by integrating the distribution. Our NLO+NNLL predictions are very stable with respect to a choice of the central scale $\mu_0$ for the invariant mass distribution, and consequently also for the total cross section.  As this is not the case for the NLO predictions, the NNLL corrections vary in size, depending on the choice of the scale. In general, for the energies and scale choices considered they provide a non-negative modification of the cross section, which for the scale choice of $\mu_0=Q$ can be even higher than 20\% at larger values of $Q^2$.

We estimate the theoretical error due to scale variation by using the 7-point method, allowing for independent variation of renormalization and factorization scales. The overall size of the theoretical scale error becomes gradually smaller as the accuracy of resummation increases, albeit the reduction is relatively modest. The reduction would have been much more significant if the scale error had been estimated by simultaneous variation of renormalization and factorization scales, i.e.\ of $\mu=\muf=\mur$. However, as it seems that the reduction in this case is a result of cancellations between factorization and renormalization scale dependencies, we choose a more conservative 7-point approach for estimating the error.

The stability of NLO+NNLL results w.r.t.\ the scale choice allows us to derive our best prediction for the $pp \to \ttbh$ total cross section at 13 TeV
$$
\sigma_{\rm NLO+NNLL}=500 ^{+7.5 \% + 3.0\%}_{-7.1\% -3.0\%} \ {\rm fb},
$$
and at 14 TeV
$$
\sigma_{\rm NLO+NNLL}=604 ^{+7.6 \% + 2.9\%}_{-7.1\% -2.9\%} \ {\rm fb},
$$
where the first error is the theoretical uncertainty due to scale variation and the second error is the pdf uncertainty. We note that the predictions are very close in their central value to the corresponding NLO predictions obtained for the scale choice $\mu_0=M/2$ and are compatible with them within errors, vindicating the appropriatness of this commonly made choice. However, in comparison with the NLO predictions obtained in this way, our NLO+NNLL predictions are characterized by the overall smaller size of the theory error related to scale variation. For an equivalent scale choice setup, our NLO+NNLL results for the $\ttbh$ production process at the LHC agree with the results previously obtained by Broggio et al.~\cite{Broggio:2016lfj}. \\

\section*{Acknowledgments}

We are grateful to M. Kr\"amer for providing us with a numerical code for  NLO $\ttbh$ cross section calculations~\cite{Beenakker:2001rj}. We would like to thank Daniel Schwartl\"ander for his input in the later stages of this work. AK gratefully acknowledges valuable exchanges with A. Kardos and Z. Trocsanyi.  This work has been supported in part by the DFG grant KU 3103/1. Support of the Polish National Science Centre grant no.\ DEC-2014/13/B/ST2/02486 is kindly acknowledged. This work was also partially supported by the U.S.\ National Science Foundation, under grants PHY-0969510, the LHC Theory Initiative, PHY-1417317 and PHY-1619867. AK would like to thank the Theory Group at CERN, where part of this work was carried out, for its kind hospitality. TS acknowledges support in the form of a Westf\"alische Wilhelms-Universit\"at Internationalisation scholarship.

\bibliographystyle{JHEP}
\providecommand{\href}[2]{#2}\begingroup\raggedright\endgroup

\end{document}